\documentclass[aps,fleqn,twocolumn,showpacs]{revtex4-1}

\usepackage{graphicx}
\usepackage{amsmath} 
\usepackage{grffile}

\def\i{{\rm i}}
\def\e#1{{\rm e}^{#1}}
\def\d{{\rm d}}

\def\len{{\rm len}}
\def\vel{{\rm vel}}
\def\velx{{\rm red}}

\def\vec#1{\mathbf{#1}}
\def\vr{\vec{r}}
\def\vp{\vec{p}}

\def\vA{\boldsymbol{{\cal A}}}

\def\Epond{E_{\rm p}}

\DeclareMathOperator*{\SumInt}{%
\mathchoice%
 {\ooalign{$\displaystyle\sum$\cr\hidewidth$\displaystyle\int$\hidewidth\cr}}
 {\ooalign{\raisebox{.14\height}{\scalebox{.7}{$\textstyle\sum$}}\cr\hidewidth$\textstyle\int$\hidewidth\cr}}
 {\ooalign{\raisebox{.2\height}{\scalebox{.6}{$\scriptstyle\sum$}}\cr$\scriptstyle\int$\cr}}
 {\ooalign{\raisebox{.2\height}{\scalebox{.6}{$\scriptstyle\sum$}}\cr$\scriptstyle\int$\cr}}
}

\def\secsuppl#1{Sect.\,#1 of the supplement \cite{suppl}}

\begin{document}
\title{Recovery of dynamic interference}
\author{Mehrdad Baghery}
\author{Ulf Saalmann}
\author{Jan M. Rost}
\affiliation{Max-Planck-Institut f{\"u}r Physik komplexer Systeme,
 N{\"o}thnitzer Str.\ 38, 01187 Dresden, Germany}
\date{\today}

\begin{abstract}\noindent
We develop general quantitative criteria for dynamic interference, a manifestation of double-slit interference in time which should be realizable with brilliant state-of-the-art high-frequency laser sources. 
Our analysis reveals that the observation of dynamic interference hinges upon maximizing the difference between the dynamic polarization of the initial bound and the final continuum states of the electron during the light pulse, while keeping depletion of the initial state small. 
Confirmed by numerical results, we predict that this is impossible for the hydrogen ground-state but feasible with excited states explicitly exemplified with the hydrogen 2p-state.
\end{abstract}


\maketitle

Interference is a basic concept ruling optical as well as quantum mechanical wave phenomena, most prominently realized through variations of the double-slit scenario.
The advent of intense laser pulses with finite pulse length has contributed a new natural double-slit scenario in the time-domain: A wave packet that is launched for some dynamic reason at a certain time of the raising part of the pulse, in principle encounters the same laser envelope amplitude
 at a certain time during the falling part of the pulse, constituting a double slit in time. If the source for the wave packet has not changed between the two ``slits'', interference of both wave packets with maximal possible contrast results, depending on the time-interval between the slits. This scenario was realized early on
 with picosecond pulses and Rydberg excitations on comparatively slow time scales in \cite{stdu+93}. It was touched upon in the context of stabilization study with high-frequency Floquet theory
for above-threshold ionization \cite{toto+07,toto+08} to finally become topical under the name dynamic interference in the soft Xray domain for femtosecond pulses \cite{dece12,dece13}. Indeed, the breathtaking development of intense light sources towards attosecond pulse lengths \cite{kriv09} and Xray frequencies \cite{mcth10} has tremendously broadened the parameter range available for light-matter interaction, and consequently for the fundamental phenomenon of dynamic interference.
 
In order to trigger experiments, and gain an understanding of the general phenomenon of dynamic interference,
in the following we will work out the parameter windows where dynamic interference is prominent on very different scales of time and energy. Formulating the relevant properties of the laser pulse and the target electron leads us to the appropriate theoretical framework for dynamic interference. Making use of the minimal analytical model described before \cite{dece13}, we will show that only its version in (reduced) velocity gauge can be safely used. 
As it turns out, the same is true for numerical implementations of dynamic interference, although for different reasons. Surprisingly and in contrast to previous claims, we also find that ionization of hydrogen from its ground-state does not exhibit dynamic interference, whereas ionization from an excited state does indeed result in dynamic interference.

The soft Xray regime we will be mainly concerned with here (electron excess energies below 100\,eV) is challenging from a theoretical point of view, since single-photon ionization in the VUV regime \emph{cannot\/} be taken as the indication for a standard perturbative light-matter coupling: Firstly, there may be substantial depletion of the ionized state during the pulse, and secondly, multi-photon processes can be involved as indicated by appreciable dynamic Stark shifts (also referred to as \textsl{ac Stark shifts}) of energies. However, due to the weak transitions in the continuum, 
multi-photon interaction does not lead to substantial multi-photon ionization in contrast to infrared or optical pulses \cite{joky+11}.

For this intermediate regime of light-matter coupling, which is neither fully perturbative nor does it lead to multi-photon ionization, we will identify the two dimensionless parameters $\delta$ and $\gamma$
accounting for the dynamic Stark shift and depletion of the initial state, respectively. 
The appearance of dynamic interference depends on a suitable ratio of these two parameters. 

We start from the standard minimal-coupling Hamiltonian written in the velocity gauge
\begin{subequations}
\label{eq:velgauge}
\begin{equation}\label{eq:hamvel}
\hat{H}^{\vel} =\frac{1}{2}\big[\hat{\vp}+\vA(t)\big]^{2}+V(\hat{\vr}),
\end{equation}
where $V$ is some external potential and 
\begin{equation}\label{eq:pulse}
\vA(t)=\vA_{0}\,g(t)\cos(\omega t)
\end{equation}
\end{subequations}
is the vector potential of the laser pulse with a Gaussian envelope $g(t) = \exp(-t^{2}/T^{2})$. 
We use atomic units throughout the text unless noted otherwise.
We represent solutions $|\psi(t)\rangle$ of the time-dependent Schr\"odinger equation for \eqref{eq:velgauge} with an expansion into field-free (bound and continuum) states $\varphi_{\alpha}$
\begin{equation}\label{eq:velEF}
\big|\psi(t)\big> =\e{-\frac{\i}{2}\!\int^{t}\!\d t'{\cal A}^{2}(t')}\SumInt_{\alpha}\big|\varphi_{\alpha}\big>\,\e{-\i E_{\alpha}t}\,a_{\alpha}(t),
\end{equation}
where the exponential pre-factor transforms away $\frac{1}{2}\vA^{2}(t)$ appearing in \eqref{eq:hamvel}. We will call this new gauge \emph{reduced\/} velocity gauge. 
The index $\alpha$ comprises all quantum numbers defining the eigenstate, which in case of continuum states are the energy $E$ and the symmetry $\kappa$. 
Standard 1st-order time-dependent perturbation theory predicts the amplitude 
(of continuum states at energy $E$)
\begin{equation}\label{eq:integ}
a_{E,\kappa} =-\i\,\vec{p}_{E,\kappa}\int\!\d t\,\boldsymbol{{\cal A}}(t)\e{\i[E-E_{\rm in}]t}
a_{\rm in}(t)\,
\end{equation}
for ionization to a final energy $E$ from initial energy $E_{\rm in}$. The dipole matrix elements $\mathbf{p}_{E,\kappa}\equiv\langle\varphi_{E,\kappa}|\hat{\vp}|\varphi_{\rm in}\rangle$ connect the initial state $\varphi_{\rm in}$ to continuum states $\varphi_{E,\kappa}$ of energy $E$. 
Due to selection rules only some of the matrix elements are non-zero.
For the photo-effect, implying weak perturbations $\vA(t)$, 
Eq.\,\eqref{eq:integ} allows for an explicit solution since one may assume that $a_{\rm in}(t)=1$ for all times.

For dynamic interference, however, the dynamic Stark shift and the depletion of the initial state become relevant.
As long as the laser envelope varies slowly compared to the laser cycle, the system remains in an adiabatic regime
where one may average the response of the system to the laser field over the laser cycle to arrive at a formulation solely expressed in terms of the laser envelope $g(t)$. 
Incorporated into \eqref{eq:integ} one obtains a modified coefficient $a_{\rm in}(t)$ which still allows 
for a solution (at least in terms of a stationary-phase approximation) as before \cite{dece12,dece13}. Hereby, the phase of 
$a_\mathrm{in}$ becomes time-dependent
\begin{subequations}\label{eq:model}\begin{align}
a_{\rm in}(t)&=\e{-\i\phi_{\rm in}^{\delta\gamma}(t)}
\\
\phi_{\rm in}^{\delta\gamma}(t)& = [\delta- \i\gamma/2]\,\Epond T\,G(t)\,,
\intertext{with the ponderomotive energy $\Epond$ and the dimensionless function $G$ defined by pulse parameters}
\label{eq:epond}
\Epond&\equiv\frac{\vA_{0}{\!}^{2}}{4}= \frac{I}{4\omega^{2}},
\\
G(t) & \equiv\frac{1}{T}\int^{t}\!\!\d t'\, g^{2}(t').
\end{align}\end{subequations}
The derivative ${\rm d}\phi^{\delta\gamma}_{\rm in}/{\rm d}t$ can be interpreted as the complex, frequency-dependent energy of the initial state in the laser pulse, proportional to the (peak) ponderomotive energy $\Epond$. Thereby, $\delta$ accounts for the Stark shift $\Delta$ (the Stark shift is indeed the time derivative of the phase $\Delta(t)=\delta\,\Epond T\,\frac{\d}{\d t}G(t)=\delta\,\Epond(t)$ with $\Epond(t)=\Epond\,g^{2}(t)$), and the decay width $\gamma$ accounts for the depletion.
Both constants depend on the laser frequency $\omega$ and can be derived 
from 2nd-order time-independent perturbation theory (see \secsuppl{1}) or equally extracted from a numerical propagation \cite{dece13}. Obviously the minimal description \eqref{eq:model} is only valid as long as Stark shift and decay width are linear in $\Epond$.

\begin{figure}[tb]
\includegraphics[width=\columnwidth]{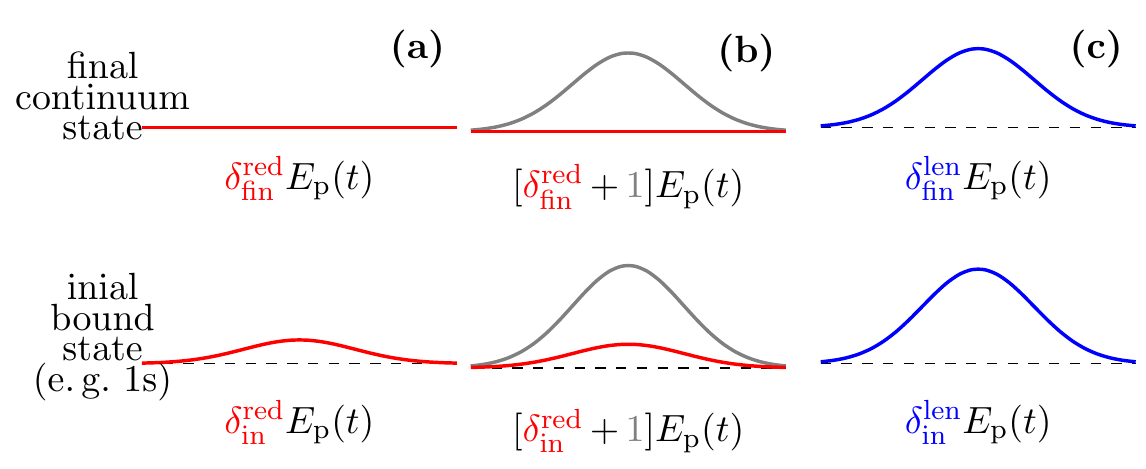}
\caption{Sketch showing the relation of the dynamic Stark shifts $\Delta(t)$ of both the initial bound and final continuum states for 
\textbf{a)} the reduced velocity gauge $\hat{H}^{\velx} {=} \frac{1}{2}\hat{\vp}^2{+}\vA(t)\hat{\vp}{+}V(\hat{\vr})$,
\textbf{b)} the velocity gauge $\hat{H}^{\vel} {=} \frac{1}{2}[\hat{\vp}{+}\vA(t)]^{2}{+}V(\hat{\vr})$, and
\textbf{c)} the length gauge $\hat{H}^{\len} {=} \frac{1}{2}\hat{\vp}^2{+}V(\hat{\vr}){-}\frac{{\rm d}}{{\rm d}t}\vA(t){\cdot}\hat{\vr}$,
respectively.
See also \secsuppl{2}.}
\label{fig:sketch1}
\end{figure}

In Eq.\,\eqref{eq:model} the dynamic Stark shift has only been introduced to the initial state, but not to the final state in the continuum. 
This is only legitimate in the reduced velocity gauge, where each state has just an intrinsic dynamic Stark shift as shown in Fig.\,\ref{fig:sketch1}a, with the one for the continuum being in general negligible.
This also applies to the velocity gauge (Fig.\,\ref{fig:sketch1}b) where all states have an \emph{additive\/} ponderomotive shift, which can be easily removed by a global phase in the wavefunction as done in Eq.\,\eqref{eq:velEF}.
In contrast, in the length gauge the ponderomotive shift \emph{cannot\/} be easily separated as indicated in Fig.\,\ref{fig:sketch1}c. 
There, to a good approximation \cite{dekr00}, the Stark shift of continuum electrons is given by the ponderomotive energy. 
Consequently, the intrinsic Stark shift of any continuum state in reduced velocity gauge is rather small and negligible, rendering the description with Eq.\,\eqref{eq:model} adequate. 

As just emphasized, in length gauge the (trivial) ponderomotive shift cannot be split off the initial or final states and therefore has to be covered by any convergent numerical calculation, which is typically much more demanding than in velocity gauge (e.\,g.\ many more partial waves are required).
This is the reason why it has been noted in long-wavelength strong-field physics that the velocity gauge is preferable for numerical calculations \cite{cola96,mu99}. 
 
From Fig.\,\ref{fig:sketch1} one obtains an intuitive understanding regarding the mechanism behind dynamic interference independent of the gauge: the initial-state energy increased by $\omega$ for single-photon absorption may intersect the energy $E$ of the final state
at two points (in time): These time instants are stationary-phase points where the time-derivative of the phase in the integral \eqref{eq:integ} vanishes, hence the amplitudes at these two dominate the integral constituting the
two-slit scenario. Dynamic interference will be most pronounced if the Stark shifts of initial and final states are very different. Quantitative details as well as the possibility for dynamic interference in the first place depend of course on the parameters entering the phase,
namely the electronic response properties $\delta(\omega)$ and $\gamma(\omega)$ in connection with the pulse properties $\Epond$ and $T$, which we will analyze next.

\begin{figure}[t]
\includegraphics[scale=0.5]{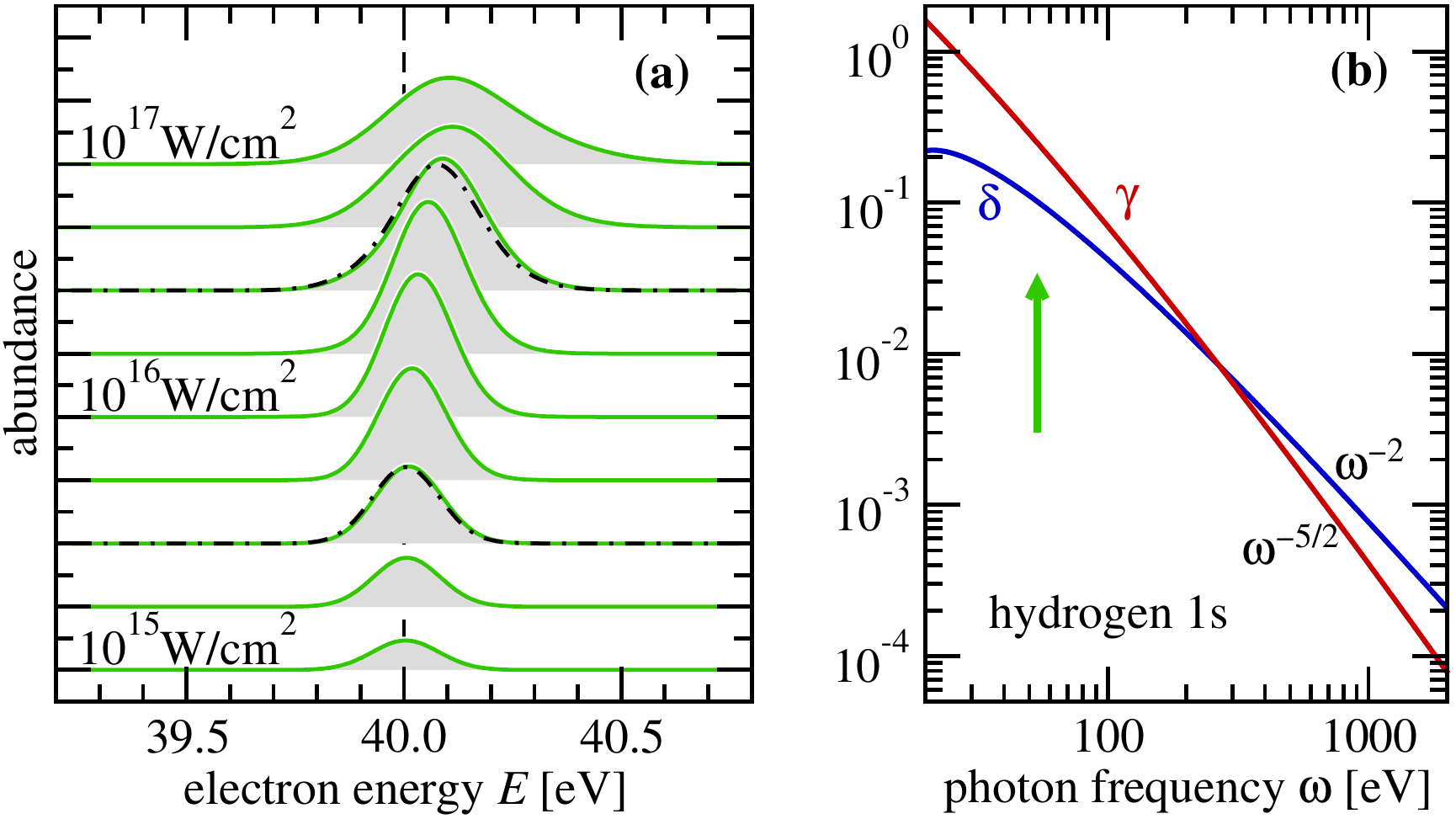}
\caption{\textbf{(a)}
Photo-electron spectra for 1s-hydrogen exposed to 10\,fs pulses with a carrier frequency of $\omega=53.60$\,eV
for intensities $I_{k}=10^{k/4}{\times}\,10^{15}$W/cm$^{2}$ with $k=0,1,\ldots,8$.
The dashed line marks the energy $E_{\omega}=E_{\rm 1s}{+}\omega=40$\,eV.
The result of the minimal model \eqref{eq:model} is shown for two intensities by dot-dashed lines.
\textbf{(b)}
Dimensionless parameters $\delta(\omega)$ and $\gamma(\omega)$ for the hydrogen 1s-state as introduced in Eq.\,\eqref{eq:model} and defined in detail \secsuppl{I}.
The asymptotic behavior for $\omega\,{\to}\,\infty$ is given for both.
The green arrow marks the frequency $\omega=53.6$\,eV used in the left panel and in previous publications \cite{dece12,dece13}.
}
\label{fig:1s-data}
\end{figure}

From the minimal model \eqref{eq:model} it is easy to to see that two conditions must be fullfilled for dynamic interference:
(i) the Stark shift must be larger than the bandwidth 
of the pulse with length $T$ in order to be energetically resolved,
(ii) depletion should be sufficiently weak in order to have ionization in the rising \emph{and\/} falling wings of the pulse.
In order to quantify these conditions we note that the bandwidth of the pulse \eqref{eq:pulse} is $\sqrt{2}/T$ and that $G(0)=\sqrt{\pi/2}$.
Thus, on one hand condition (i) is satisfied if
\begin{equation}\label{eq:cond2}
\delta \Epond>\sqrt{2}/T
\quad\mbox{or}\quad
\Epond T>\frac{\sqrt{2}}{\delta}.
\end{equation}
On the other hand, condition (ii) is fulfilled if
\begin{equation}\label{eq:cond1}
\frac{\gamma}{2}\Epond T\,G(0)<1
\quad\mbox{or}\quad
\Epond\,T<\frac{\sqrt{2/\pi}}{\gamma}.
\end{equation}
These two conditions give lower and upper limits for the product $\Epond T$. 
Apparently they can be met simultaneously\,---\,thus allowing for dynamic interference\,---\,only if
\begin{equation}\label{eq:optet}
\delta>\sqrt{\pi}\gamma,
\end{equation}
which implies that in the competition between Stark shift and depletion the former should dominate.
This condition holds for any atom or molecule.
As a consequence, we can predict the laser parameters for which one will observe dynamic interference, provided the response parameters $\delta(\omega)$ and $\gamma(\omega)$ are known. They are shown 
in Fig.\,\ref{fig:1s-data}b for the ground-state of hydrogen as an example.

\def\omegaequal{\widetilde{\omega}}
Having condition \eqref{eq:optet} in mind, one sees from Fig.\,\ref{fig:1s-data}b that this requires frequencies larger than $\omegaequal\,{=}\,265$\,eV where $\gamma(\omegaequal)=\delta(\omegaequal)$. For $\omega\,{\gg}\,\omegaequal$ the absolute values are very small, hence it is virtually impossible to observe dynamic interference for the ground-state of hydrogen. This is confirmed by the numerical photo-absorption spectra in Fig.\,\ref{fig:1s-data}a determined by direct propagation of the time-dependent Schr\"odinger equation in (reduced) velocity gauge for the photon frequency $\omega$ marked by the green arrow in Fig.\,\ref{fig:1s-data}b.
Numerical details are given in \secsuppl{3}, the parameters used are $\ell_{\rm max}{=}4$, $r_{\rm max}{=}3000$\,a$_{0}$, $n{=}3000$ and $E_{\rm max}{\approx}134$\,eV.
As one can see from Fig.\,\ref{fig:1s-data}a, the spectrum has a single photo-electron peak which gets Stark-shifted and broadened for increasing intensities while keeping the pulse length fixed at $T{=}10$\,fs. 
The results have been confirmed with two other packages \cite{bako06,pamu16}
for the numerical propagation of the time-dependent Schr\"odinger equation.  

\begin{figure*}[t!]
\includegraphics[scale=0.5]{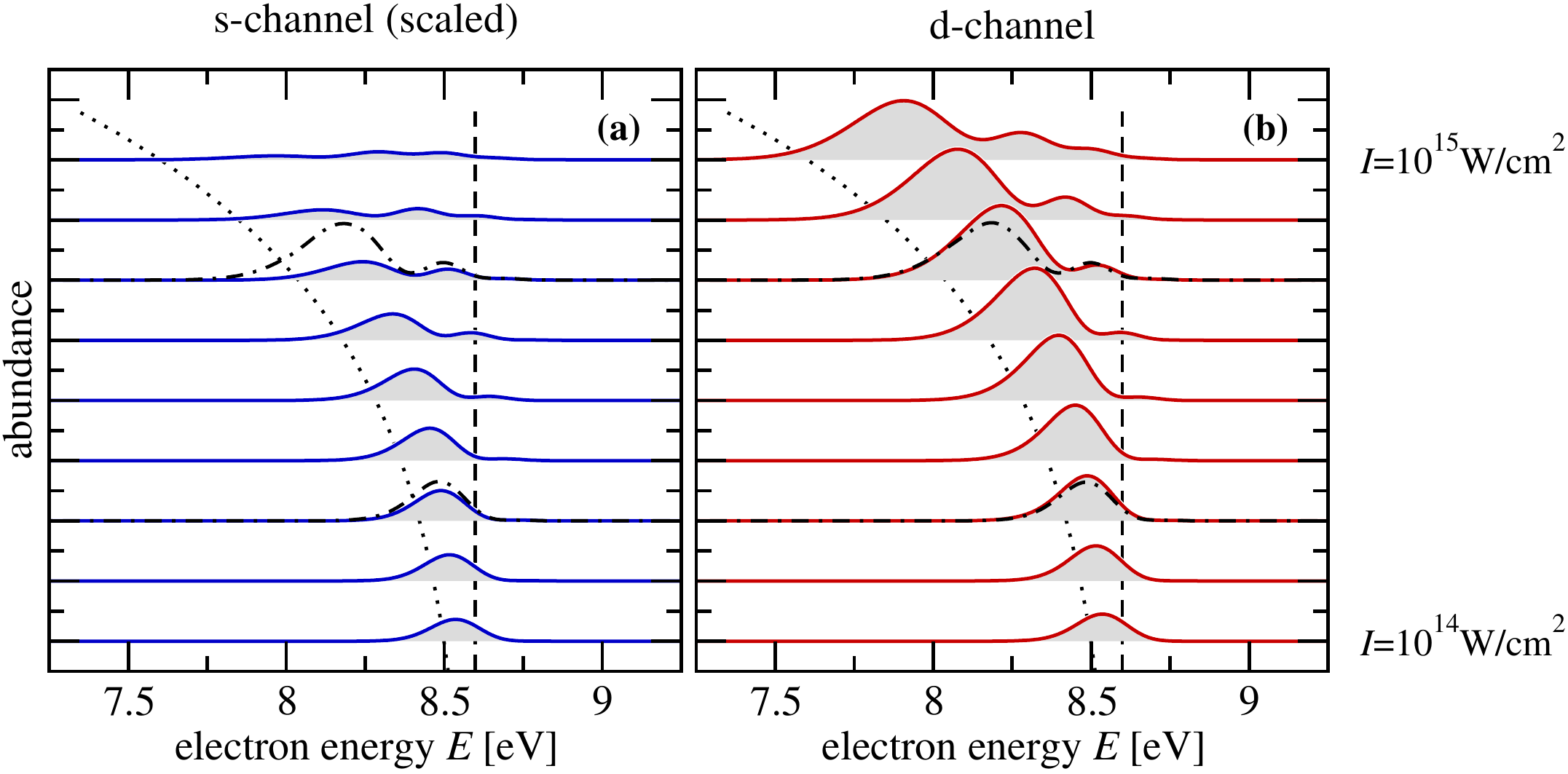}\hfill
\includegraphics[width=0.7\columnwidth]{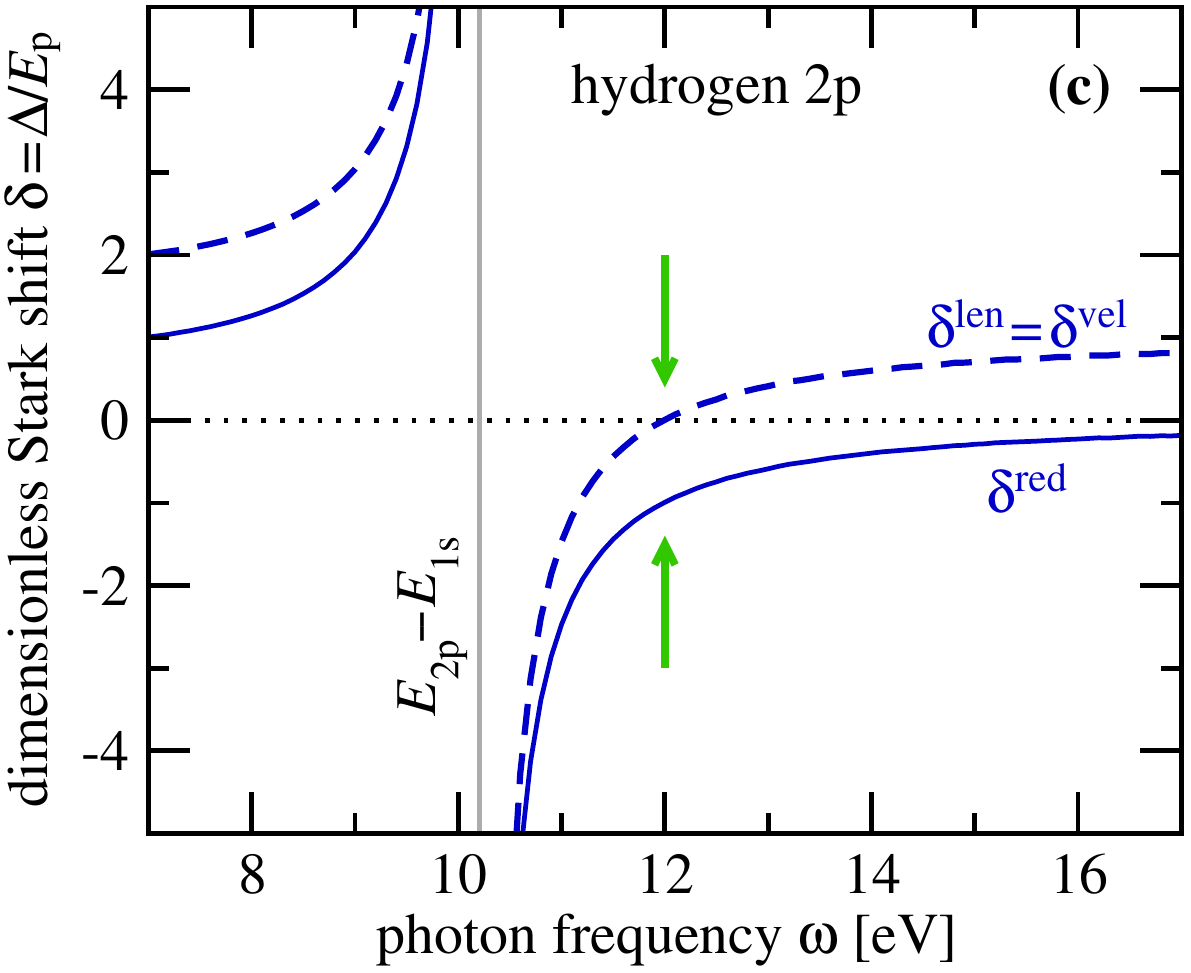}
\caption{Photo-electron spectra for 10\,fs pulses with a carrier frequency of $\omega=12$\,eV for hydrogen in the 2p-state.
Ionization into the s-channel \textbf{(a)} and the d-channel \textbf{(b)} is shown.
The intensities from bottom to top are $I_{k}=10^{k/8}{\times}\,10^{14}$W/cm$^{2}$ with $k=0,1,\ldots,8$.
The dashed line marks the energy $E_{\omega}=E_{\rm 2p}+\omega=8.59875$\,eV, the dotted line interpolates between the energies $E_{\omega,k}=E_{\rm 2p}+\omega-\Epond(I_{k})$.
The s-channel is scaled by the factor $f$ in order to make the height of the peak equal in the perturbative limit: $f=|\vp_{E_{\omega},{\rm s}}|/|\vp_{E_{\omega},{\rm d}}|\approx6.137$.
The result of the minimal model \eqref{eq:model} is shown for two intensities by dot-dashed lines.
\textbf{(c)}
Dimensionless Stark shift $\delta$ for the 2p state of hydrogen as a function of the photon frequency $\omega$ in units of the ponderomotive energy $\delta\equiv\Delta/\Epond$.
We show $\delta^{\len}{=}\,\delta^{\vel}$ (blue-solid), and $\delta^{\velx}=\delta^{\vel}-1$ (blue-dashed),
cf.\ \secsuppl{2}.
The green arrow marks the photon frequency $\omega=12$\,eV, where the Stark shift vanishes and which was used in the two left panels.
The Stark shift diverges at the transition energy $E_{\rm 2p}-E_{\rm 1s}$ (gray-solid).
}
\label{fig:2p-data}
\end{figure*}%

In contradiction to these results, dynamic interference has been reported for the hydrogen ground-state \cite{dece12,dece13,deho+13,yufu+13}. 
Even more puzzling, the numerical findings (obtained in length gauge) seem to be supported by corresponding results from the minimal model in length gauge appearing in the same papers. It is very rare and unfortunate that two independent mistakes, both related to a faulty usage of the length gauge, one in the model and one in the numerical calculation, should lead to agreeing results, seemingly re-assuring the findings regarding dynamic interference.
 
The analytical mistake is easy to identify and originates from using the minimal model \eqref{eq:model} in the length gauge but leaving out the Stark shift in the continuum state (see Fig.\,\ref{fig:sketch1}c). Thereby, the difference between the time-dependent energies of initial and final states are artificially increased, leading to the appearance of stationary-phase points of the integrand in \eqref{eq:model} at the wrong intensities. The numerical calculations were carried out in length gauge with a limited number of partial waves, not enough to properly describe the ponderomotive motion of the electron. 
This parallels the mistake in the analytical model, where the ponderomotive energy dependence was left out and therefore lead to the accidental agreement of numerical and analytical calculations in those papers \cite{dece12,dece13,deho+13,yufu+13}.
A detailed comparison of numerical calculations in length and velocity gauges for different maximal angular momenta can be found in \secsuppl{3}.
 
Extrapolating from the conditions for dynamic interference in ionizing hydrogen, it seems very difficult to 
realize this phenomenon with state-of-the art laser systems. Yet, this can be easily achieved starting from the 2p excited state, which at the same time highlights the relevance of considering the proper Stark shifts. 
Starting from 2p and choosing appropriate intensities and frequencies of the laser, we can prepare an effective initial state which has no Stark shift in length gauge (or a large negative Stark shift in reduced velocity gauge).
This implies a large difference in the Stark shifts between continuum and initial bound states and therefore offers excellent conditions for dynamic interference.

At a frequency of $\omega=12$\,eV the dynamic Stark shift of the 2p state vanishes (in length and velocity gauge) and is therefore given by $\Delta(t)=-\Epond(t)$ in reduced velocity gauge, as can be seen in Fig.\,\ref{fig:2p-data}c.
At this frequency the coupling to the 1s-state fully compensates the coupling to all other states such that the polarizability and the Stark shift vanish. We have performed a propagation with the same parameters as before, and obtained the spectra shown in Figs.\,\ref{fig:2p-data}a and \ref{fig:2p-data}b, cf.\ details in \secsuppl{2}.
Since we start from a p-state, photo electrons are emitted into s- and d-channels, where the yield for the latter is larger due to the larger dipole matrix element.
However, qualitatively both angular momentum channels exhibit the same behavior for increasing intensities $I$.
For low intensities the spectrum is Gaussian-shaped and slightly red-shifted with respect to $E_{\omega}=E_{\rm 2p}+\omega$. This shift increases with larger $I$ and for $I\gtrsim5{\times}\,10^{14}$W/cm$^{2}$ one clearly sees dynamic interference in both channels. We note in passing that according to earlier publications \cite{dece12,dece13} one would not expect any dynamic interference at all here.
Moreover, in contrast to the blue shift predicted previously, we observe a red-shift increasing with intensity 
which follows directly from $\Delta(t)<0$ mentioned above. 

In summary, by formulating single-photon ionization in 1st-order time-dependent perturbation theory with phases obtained from the 2nd order, 
we have derived quantitative conditions under which dynamic interference can occur. The approach is tailored towards the dynamic regime of non-perturbative single-photon ionization, characteristic of interactions with intense soft Xrays and has allowed for the separation of the frequency-dependent response of the electronic system in terms of Stark shift $\delta(\omega)$ and depletion $\gamma(\omega)$ and the time-dependent laser pulse envelope. This separation facilitates the determination of the electronic response by electronic structure calculations and helps to accurately assess experimental conditions for dynamic interference.

\def\articletitle#1{{\it #1},}

\end{document}